\documentstyle[11pt]{article}
\begin{document}
\baselineskip 18pt
\begin{titlepage}
\centerline{\large\bf  A POSSIBLE HIDDEN SYMMETRY AND GEOMETRICAL SOURCE}

\vspace{0.2cm}

\centerline{\large\bf OF THE PHASE IN THE CKM MATRIX}

\vspace{1cm}

\centerline{Jing-Ling Chen$^1$, Mo-Lin Ge$^{1,2}$,
            Xue-Qian Li$^{3,4}$, Yong Liu$^1$}
\vspace{0.5cm}
{\small
\centerline{\bf 1. Theoretical Physics Division}
\centerline{\bf Nankai Institute of Mathematics}
\centerline{\bf Nankai University, Tianjin 300071, P.R.China}
\centerline{ Email:   yongliu@sun.nankai.edu.cn}
\centerline{ Email:   yongliu@ns.lhp.ac.cn}
\centerline{ Fax: 0086-22-23501532, Tel: 0086-22-23501029}
\vspace{0.2cm}
\centerline{\bf 2. Center for Advanced Study }
\centerline{\bf Tsinghua University, Beijing 100084, P.R.China}
\vspace{0.2cm}
\centerline{\bf 3. CCAST(World Laboratory) P.O.Box 8730, Beijing 100080,
                P.R.China}
\vspace{0.2cm}
\centerline{\bf 4. Department of Physics }
\centerline{\bf Nankai University, Tianjin 300071, P.R.China}
}
\vspace{2cm}

\begin{center}
\begin{minipage}{12cm}

\noindent{\bf Abstract}

\vspace{0.2cm}

Based on the present data, the three CKM angles may construct a spherical
surface triangle whose area automatically provides a "holonomy" phase.
By assuming this geometrical phase to be that in the CKM matrix determined
by an unknown hidden symmetry, we compare the theoretical prediction on
$\epsilon$ with data and find they are consistent within error range. We also
suggest restrictions for the Wolfenstein parameters explicitly, the
agreement will be tested by more precise measurements in the future. At
least we can claim that such geometrical phase should be part of the weak
phase appearing in the CKM matrix, even not the whole.

PACS number(s): 11.30.Er, 12.10.Ck, 13.25.+m

\end{minipage}

\end{center}

\end{titlepage}

\newpage

\centerline{\bf   A Possible Hidden Symmetry and Geometrical Source}

\centerline{\bf
of The Phase in The CKM Matrix}

\vspace{1cm}

Although more than thirty years have elapsed since the discovery of CP
violation \cite{Chris}, our understanding about the source of CP violation
is still very poor. In the Minimal Standard Model (MSM), CP violation
is due to the presence of a weak phase in the Cabibbo-Kabayashi-Maskawa
(CKM) matrix \cite{KM}\cite{Cabi}. 
Up to now, all the experimental results are in good agreement with MSM.
Nevertheless, the correctness of CKM mechanism is far from being proved.
Search for the source of CP violation is a profound and hard task in
high energy physics \cite{Jarl,WolfCP,Chau,Pasch,Pich}. In this work
we  restrict ourselves in the framework of MSM and see if we can find
something which was missing in previous studies. First, let us review
what we have learned about the CKM matrix.

(1) Considering all constraints on the matrix elements, for three generations,
there are three independent angles and a weak phase which cannot be rotated
away or absorbed into the quark wavefunctions. The phase is also independent
of the three angles in principle. From another point, much effort has been
made to understand the source of the three rotation angles and the phase.

Fritzsch
\cite{FritPh,FritNu} noticed
that because the eigenstates of the weak interaction are not the quark
mass-eigenstates, there should be a unitary transformation to connect the
two bases. It would establish a certain relation between the
quark masses and the weak interaction mixing angles, while a weak CP
phase is embedded explicitly.

From the general theory of Kabayashi-Maskawa
\cite{KM},
we know that there can exist a phase factor in the three-generation
CKM matrix and it cannot be rotated away by re-defining the phases of
quarks, but we can ask whether there is an intrinsic relation between
the phase and the three rotation angles.

In Fritzsch's theory, the CKM matrix comes from diagonalizing the U-type and
D-type quark mass matrices as it is possible that there are certain
horizontal relations between different generations of quarks. The proposed
symmtry has undergone some modifications for fitting data, especially
the top-quark mass. These horizontal relations which determine the off-diagonal
elements of the mass matrices which are proportional to a typical quantity
$\lambda$. It may hint that there is a broken horizontal symmetry and
the scale of breaking is related to $\lambda$.

Concretely, supposing $V_d$ and $V_u$ diagonalize the mass matrices for
d-type and u-type quarks respectively \cite{Don}, $V_{KM}
\equiv V_u^{\dagger}V_d$
is the CKM matrix and can be written as
\begin{equation}
V_{KM}= \left (
\begin{array}{ccc}
   c_1 & -s_1c_3& -s_1s_3 \\
   s_1c_2 & c_1c_2c_3-s_2s_3e^{i\delta}& c_1c_2s_3+s_2c_3e^{i\delta}\\
   s_1s_2 & c_1s_2c_3+c_2s_3e^{i\delta}& c_1s_2s_3-c_2c_3e^{i\delta}
\end{array}
\right )
\end{equation}
with the standard notations $s_i=\sin\theta_i$ and $c_i=\cos\theta_i$.

It is noted that here we adopt the original form of the CKM parametrization.
There are some other parametrization ways, for example the Wolfenstein's
\cite{Wolfen} \cite{Zzx}\cite{Buras}
and that recommended by the data group \cite{ChK}\cite{Mai}
\cite{Plankl}, but it is believed that physics
does not change when adopting various parametrizations.

It is well known that the KM parametrization can be viewed as a product of
three Eulerian rotation matrices and a phase matrix \cite{Don}
\begin{equation}
V_{KM}= \left (
\begin{array}{ccc}
   1 & 0 & 0 \\
   0& c_2 &-s_2 \\
   0 & s_2 & c_2
\end{array}
\right )
\left (
\begin{array}{ccc}
   c_1 &-s_1 & 0 \\
   s_1& c_1 & 0 \\
   0 & 0 & 1
\end{array}
\right )
\left (
\begin{array}{ccc}
   1 &0& 0 \\
   0& 1& 0 \\
   0 & 0 & -e^{i\delta}
\end{array}
\right )
\left (
\begin{array}{ccc}
   1 &0& 0 \\
   0& c_3& s_3 \\
   0 & -s_3 & c_3
\end{array}
\right ).
\end{equation}

People have noticed that the weak CP phase $\delta$, which cannot
be eliminated in the three generation CKM matrix by any means, is introduced
artificially and seems to have nothing to do with the three "rotation"
angles. Anyway, such a fact does not seem to be natural.

As this concept is widely accepted, we may ask if there may exist
a hidden symmetry which can relate the weak phase with the three rotation
angles? In general, there is nothing to forbid it and of course nothing
to confirm it either, we are only discuss a possible source. Only one thing
we are sure is that there indeed exists a geometrical phase and it can
serve as one of the sources to the CP phase.

(2) Based on observation, the recently measured $\theta_1,\theta_2,\theta_3$
satisfy
$$\theta_i+\theta_j\geq\theta_k,\;\;\;\;\; i,j,k=1,2,3,$$
and if we only take the positive values of $\sin\theta_i$ as $0\leq\theta_i
\leq\pi/2$ (i=1,2,3), then
$$\theta_1+\theta_2+\theta_3\leq{3\pi\over 2}.$$
Therefore the three angles can construct a spherical surface triangle on a
unit sphere in the Hilbert space.

The three angles correspond to three arcs on the unit sphere and they
enclose an area $\delta$. The $\delta$ and the three angles have a definite
relation
\begin{equation}
\label{del}
\cos{\delta\over 2}={1+\cos\theta_1+\cos\theta_2+\cos\theta_3\over
4\cos{\theta_1\over 2}\cos{\theta_2\over 2}\cos{\theta_3\over 2}}.
\end{equation}

The geometrical meaning of the area is clear. The three vertices A, B and C
correspond to $\angle A,\angle B$ and $\angle C$. At each vertex, there
are two tangents along the two ajacent arcs as defining the positve directions
of arcs anti-clockwisely. If one moves one tangent $\vec t_1^A$ along the arc
$AB$ to vertex B and then it becomes to $\vec t_2^B$. Rotate $\vec t_2^B$
anti-clockwisely to $\vec t_1^B$ by an angle $\pi-\angle B$, then let it move
to vertex C along arc $BC$ and rotate $\vec t_2^C$ to $\vec t_1^C$, finally
move it back to vertex A and the resultant $\vec t_2^A$ which spans an angle
$\pi-\angle A$ with the original vector $\vec t_1^A$.  Geometrically,
the three angles $\alpha_1, \alpha_2, \alpha_3$ which transform $\vec t_2^{
A,B,C}$ to $\vec t_1^{A,B,C}$ respectively have the relation
\begin{eqnarray}
\alpha_1+\alpha_2+\alpha_3 &=& \pi-\angle A+\pi-\angle B+\pi-\angle C
=3\pi-(\angle A+\angle B+\angle C)\\
&=& 3\pi-(\pi+\delta)=2\pi-\delta,
\end{eqnarray}
where $\delta$ is exactly that area enclosed by the three arcs and is
called the "angular excess". It is noted that for a planar triangle,
there is no such "angular excess", so the $\delta$-phase is obviously
caused by the curved space characteristics, ie.e the affine connection.

Thus $\delta$ represents the tagent transformation along the spherical
surface triangle, so it can be the variable of an U(1) holonomy
transformation group. So it automatically corresponds to a phase $e^{i
\delta}$, which is a geometrical phase.

The three Euler angles $\theta_{12},\theta_{23},\theta_{31}$ bridge the
three generation quark flavors and the adopted  $\theta_1,
\theta_2,\theta_3$ are nothing new but an alternative parametrization
scheme, so the U(1) phase $e^{i\delta}$ would appear in the CKM matrix
afterwards.

As well-known, for a naive $O(3)$ rotation group, a geometric phase can
automatically arise while two non-uniaxial successive rotation
transformations being performed \cite{Wilc,KMSM,Jord,Simon,Wein,Arav}. For
instance, $R_x(\theta_1)$ denotes a clockwise rotation about the x-axis
by $\theta_1$, while $R_y(\theta_2)$ is about the y-axis by $\theta_2$.
Supposing on a unit-sphere surface, the positive z-axis intersects with
the surface at a point A, after performing these two sequential operations
$R_y(\theta_2)R_x(\theta_1)$, the point-A would reach point-B via an
intermediate point-C, by contrast, one can connect A and B by a single
rotation $R_{\hat \xi}(\theta_3)$, where $R_{\hat \xi}(\theta_3)$ denotes a
clockwise rotation about the ${\hat \xi}-$axis by $\theta_3$. The geometric
meaning can be depicted in a more obvious way is that if one chooses an
arbitrary tangent vector
$\hat \alpha$ at point-A which would rotate to $\hat \alpha'$ and $\hat
\alpha \prime \prime$ by $R_y(\theta_2)
R_x(\theta_1)$ and $R_{\hat \xi}(\theta_3)$ respectively, then one can find
that $\hat \alpha'$ does not coincide with $\hat \alpha \prime \prime$, but
deviates by an extra rotation. Concretely, if one writes down the rotation
in the adjoint representation of $O(3)$, he can find
\begin{equation}
\label{xi}
R_{\hat \eta}(\delta)R_{\hat \xi}(\theta_3)=R_y(\theta_2)R_x(\theta_1),
\end{equation}
where $R_{\hat \eta}(\delta)$ represents a counterclockwise rotation about
the ${\hat \eta}-$axis by $\delta$. This resultant phase is obviously
non-removable.\\

(3) Assuming that this geometrical U(1) phase is the weak phase in the CKM
matrix, namely, it means that there is no any underlying physical principle to
cause the weak phase in the matrix, but only this holonomy phase plays
role of the weak phase.  Thus we can find some consequential deductions
which would be tested by its phenomenological applications in comparison
with corresponding experimental data.

(i) A test from $\epsilon$ in $K^0-\bar K^0$ system.

So far the only reliably measured CP violation quantity is $\epsilon$ in
the K-system and the mechanism causing $K^0-\bar K^0$ mixing has already
been well studied in the framework of MSM. Except an unknown B-factor,
one can evaluate $\epsilon$ in terms of the CP phase $\delta$ as
\cite{Cheng}\cite{EGN}\cite{EGNR}
\begin{equation}
\label{CP}
|\epsilon|\approx \cos\theta_2\sin\theta_2\sin\theta_3\sin\delta
\left[{\sin^2\theta_2(1+\eta\log\eta)-\cos^2\theta_2\eta(1+\log\eta)
\over\sin^4\theta_2+\cos^4\theta_2\eta-2\sin^2\theta_2\cos^2\theta_2\eta\log
\eta}\right],
\end{equation}
where $\eta=m_c^2/m_t^2$.

The inputs of $\mid V_{ij} \mid $ are taken from the data book
\cite{PDG} and
$$m_c=1.5 \;{\rm GeV},\;\;\; m_t(m_t^2)=176\; {\rm GeV},\;\;\; |\epsilon|=
2.3\times 10^{-3},$$
with all the given errors.

By using eq.(\ref{del}) we obtain
$$\delta=0.01138\pm 0.0049,$$
while extracting the $\delta- $value from eq.(\ref{CP}), it is
$$\delta=0.0044\pm 0.0030.$$

Therefore, one can notice that considering the experimental error tolerance,
the two obtained values are roughly consistent. Since the extraction of
$\delta$ from the data $\epsilon$ still depends on the evaluations of concerned
hadronic transition matrix elements which are not reliable so far,
namely, we cannot handle the non-perturbative QCD effects satisfactorily, the
deviation between two $\delta-$values is reasonable and tolerable.

Anyhow, this phenomenological application of eq.(\ref{del}) does not
contradict to the data, in other words, may obtain some sort of support
from this comparison.

(ii) The ranges of the Wolfenstein's parameters.

Now, let us have a look at the possible ranges of the Wolfenstein's parameters
\cite{Wolfen} if the assumption that the $\delta$ given in eq.(\ref{del})
is the weak phase in the CKM matrix, is valid. In the Wolfenstein's
parametrization, the CKM matrix reads as
\begin{equation}
V_{W}= \left (
\begin{array}{ccc}
   1-\frac{1}{2}\lambda^2 & \lambda & A\lambda^3(\rho-
   i\eta+i\eta\frac{1}{2}\lambda^2) \\
   -\lambda & 1-\frac{1}{2}\lambda^2-i\eta A^2 \lambda^4 &
   A\lambda^2(1+i\eta\lambda^2)\\
   A\lambda^3(1-\rho-i\eta) & -A\lambda^2 & 1
\end{array}
\right ).
\end{equation}
So, it is necessarily to convert the new constraint Eq.(1) which
represented by four angles $\delta,\; \theta_1,\; \theta_2,\; \theta_3$
into the one represented by Wolfenstein's parameters $A,\; \lambda,
 \; \rho,\;\eta$. It is easy to do so if we take use of the following
translation prescription between KM's and Wolfenstein's parameters [11]
\begin{equation}
   s_1\approx \lambda, \;\;\;\;  c_1\approx 1-\frac{\lambda^2}{2}
\end{equation}
\begin{equation}
   s_2\approx \lambda^2 A[(\rho-1)^2+\eta^2]
\end{equation}
\begin{equation}
   s_3\approx(\rho^2+\eta^2)^{1/2}A\lambda^2
\end{equation}
\begin{equation}
   sin\delta\approx \frac{\eta}{(\rho^2+\eta^2)^{1/2}}\frac{1}{
   [(\rho-1)^2+\eta^2]^{1/2}}.
\end{equation}

From Eq.(1), we obtain
\begin{equation}
sin\delta=\frac{(1+cos\theta_1+cos\theta_2+cos\theta_3)\sqrt{
sin^2\theta_1+sin^2\theta_2+sin^2\theta_3-2(1-
cos\theta_1 cos\theta_2 cos\theta_3)}}{
(1+cos\theta_1)(1+cos\theta_2)(1+cos\theta_3)}.
\end{equation}
Substituting Eqs.(9-11) to Eq.(13) and expanding the right hand side of Eq.(13)
in powers of $\lambda$, with a little more complicated calculation,
when approximate to the order of $\lambda^5$, we get
\begin{equation}
sin\delta=\frac{A \sqrt{(1-\rho)^2+\eta^2+(\rho^2+\eta^2)}}
{2 \sqrt{2}}\lambda^3
+\frac{A [(1-\rho)^2+\eta^2+(\rho^2+\eta^2)-2A^2(1-2 \rho)^2 ]}
{2^4\sqrt{2} \sqrt{(1-\rho)^2+\eta^2+(\rho^2+\eta^2)}}\lambda^5
\end{equation}
Identify the right hand sides of Eq.(12) and Eq.(14), we have
\begin{equation}
\frac{\eta}{(\rho^2+\eta^2)^{1/2}[(1-\rho)^2+\eta^2]^{1/2}}
\approx \frac{A \sqrt{(1-\rho)^2+\eta^2+(\rho^2+\eta^2)}}
{2 \sqrt{2}}\lambda^3.
\end{equation}
Here, in comparison with $\lambda^3$, we have neglected the
term of order $\lambda^5$.

Eq.(15) is the new constraint on CP-violation and quark-mixing
represented by Wolfenstein's parameters approximate to the order
of $\lambda^3$.

In following, we want to give a simple numerical analysis. Let
\begin{equation}
x=(\rho^2+\eta^2)^{1/2}
\end{equation}
\begin{equation}
y=[(1-\rho)^2+\eta^2]^{1/2}
\end{equation}
then
\begin{equation}
\eta=\frac{1}{2} \sqrt{2 (x^2+y^2)-(x^2-y^2)^2-1}
\end{equation}
Substituting Eqs.(16-18) to Eq.(15), we arrive
\begin{equation}
\label{rel}
\frac{\sqrt{2 (x^2+y^2)-(x^2-y^2)^2-1}}{x y}=\frac{A \lambda^3}{\sqrt{2}}
\sqrt{x^2+y^2}
\end{equation}

Fixing $\lambda=0.22$ and $A=0.808\pm 0.058$ \cite{Rosner},
if we take $y=0.54\sim 1.40$ as input, then
$0.22 \sim 0.46$ for $x$ is permitted.
Hence, we find that the results are well in agreement with the experimental
analysis \cite{Pich}
\begin{equation}
x=\sqrt{\rho^2+\eta^2}=0.34\pm 0.12
\end{equation}
and
\begin{equation}
y=\sqrt{(1-\rho)^2+\eta^2}=0.97\pm 0.43.
\end{equation}

Here we draw a graph of $x^2$ versus $y^2$ using the relation given in
eq.(\ref{rel})
and we obtain two branches of
solutions where we let A to be deviated from its central value of 0.808
by a small fractions and the curves in fact are narrow bands.

From the figure, one can see that within the experimental error ranges
of $x$ and $y$, there exist solutions. It indicates that the results
do not contradict to the CKM matrix elements measurement.\\

(4) Our conclusion and discussion.

In this work, based on observation of the measured values of the CKM matrix
elements and consideration of a possible hidden symmetry, we study a
possibility that the weak phase in the CKM matrix is due to a geometrical
reason which is fully determined by the three rotation angles.

In this assumption, the geometrical phase takes responsibility of all the
roles of the weak phase in the CKM matrix. Obviously, there can be
some physical mechanisms which can also result in the weak phase and present
data cannot eliminate this possibility at all. What we show in this work is that
the geometrical phase does exist and can be a part of the phase at the
CKM matrix, moreover the bald assumption cannot eliminate existence of other
physical sources of the weak phase at all. If this geometrical phase can be
the whole of the CKM phase or only a part of it should wait for more precise
experimental measurements in the future. If the predicted curves drawn
in Fig.1, do not have common range with the measure data, it would indicate
that the geometrical phase is only a part, no matter small or large, of the
phase in the CKM matrix. \footnote{In our previous works, hep-ph/9711330
and hep-ph/9711293, our claim on the geometrical phase went too extreme,
that we attributed existence of the phase to geometrical reason and it does
not have any proof or enough decisive support from data.}

In conclusion, we can
claim a possible source of the CKM phase due to the geometrical reason and
need to wait for future experiments to test if it is the only source or
only one of the sources, i.e. there should exist other physical sources
to cause this CKM phase.\\

\vspace{0.5cm}

\noindent {\bf Acknowledgment}: This work is partly supported by National
Natural Science Foundation of China. We would like to thank Dr. Z.Z. Xing
for helpful comments and Dr. X.G. He, Dr. Y.L. Wu for fruitful discussions.


\begin{thebibliography}{99}
\bibitem{Chris} J.H.Christenson, J.W.Cronin, V.L.Fitch and R.Turlay,
            Phys.Rev.Lett. 13(1964)138.
\bibitem{KM} M.Kobayashi and T.Maskawa, Prog.Theor.Phys. 42(1973)652.
\bibitem{Cabi} N.Cabibbo, Phys.Rev.Lett. 10(1963)531.
\bibitem{Jarl} $CP\;\; Violation$ Ed. C.Jarlskog. World Scientific
            Publishing Co.Pte.Ltd 1989.
\bibitem{WolfCP} $CP \;\; Violation$ Ed. L.Wolfenstein, North-Holland,
             Elsevier Science Publishers B.V. 1989.
\bibitem{Chau} L.L.Chau, Phys.Rept. 95(1983)1.
\bibitem{Pasch} E.A.Paschos and U.Turke, Phys.Rept. 4(1989)145.
\bibitem{Pich} A.Pich, Preprint CP violation, CERN-TH.7114/93.
\bibitem{FritPh} H.Fritzsch, Phys.Lett.B. 70(1977)436.
              Phys.Lett.B. 73(1978)317.
\bibitem{FritNu} H.Fritzsch, Nucl.Phys.B. 155(1979)189.
              Preprint MPI-PHT/96-32.
\bibitem{Don} J.F.Donoghue, E.Golowich and B.R.Holstein, $Dynamics\;\;of\;\;
              the\;\;Standard\;\;Model$ Cambridge University Press, 1992.
              P.60$\sim$P.69.
\bibitem{Wolfen} L.Wolfenstein, Phys.Rev.Lett. 51(1983)1945.
\bibitem{Zzx} Z.Z.Xing, Phys.Rev.D. 51(1995)3958.
\bibitem{Buras} A.J.Buras, M.E.Ladtenbacher and G.Ostermaier,
               Phys.Rev.D. 50(1994)3433.
\bibitem{ChK} L.-L.Chau and W.-Y.Keung, Phys.Rev.Lett. 53(1984)1802.
\bibitem{Mai} L.Maiani, Phys.Lett.B 62(1976)183.
\bibitem{Plankl} H.Fritzsch and J.Plankl, Phys.Rev.D 35(1987)1732.
\bibitem{Wilc} Eds. A.Shapere and F.Wilczek, $Geometric\;\; Phases\; \;
               in\;\; Physics$, World Scientific, Singapor, 1989.
\bibitem{KMSM} G.Khanna, S.Mukhopadhyay, R.Simon and N.Mukunda,
              Ann.Phys. 253(1997)55.
\bibitem{Jord} T.F.Jordan, J.Math.Phys. 29(1988)2042.
\bibitem{Simon} R.Simon and N.Mukunda, J.Math.Phys. 30(1989)1000.
\bibitem{Wein} S.Weinberg, $The\;\;Quantum\;\;Theory\;\;of\;\;Fields$
               Published by the Press Syndicate of the University of
               Cambridge, 1995. P.81$\sim$P.100.
\bibitem{Arav} P.K.Aravind, Am.J.Phys. 65(1997)634.
\bibitem{Cheng} T.P.Cheng, L.F.Li, $Gauge\;\; Theory\;\; of\;\;
             Elementary\;\;Particle\;\; Physics$, Clarendon Press.Oxford.1984.
\bibitem{EGN} J.Ellis, M.K.Gaillard and D.V.Nanopoulos,
             Nucl.Phys.B. 106(1976)292. Nucl.Phys.B. 109(1976)216.
\bibitem{EGNR} J.Ellis, M.K.Gaillard, D.V.Nanopoulos and S.Rudaz,
             Nucl.Phys.B. 131(1977)285.
\bibitem{PDG} Particle Data Group, Phys.Rev.D. 54(1996)94.
\bibitem{Rosner} J. Rosner, hep-ph/9610222.
\end{thebibliography}
\end{document}